\documentclass[aps,pra,preprint,superscriptaddress]{revtex4}
\usepackage{graphicx}% Include figure files
\usepackage{dcolumn}% Align table columns on decimal point
\usepackage{bm}% bold math
\usepackage[usenames]{color}
\usepackage{amsmath}

\usepackage{amssymb}

\begin{document}

\preprint{preprint}

\title{Aharonov-Bohm Effect at liquid-nitrogen temperature:\\
Fr\"ohlich superconducting quantum device
}

\author{M. Tsubota}
\affiliation{Department of Applied Physics, Hokkaido University,
Sapporo 060-8628, Japan}

\author{K. Inagaki}
\affiliation{Department of Applied Physics, Hokkaido University,
Sapporo 060-8628, Japan}
\affiliation{Center of Education and Research for Topological Science and Technology, Hokkaido University,
Sapporo 060-8628, Japan}

\author{S. Tanda}
\affiliation{Department of Applied Physics, Hokkaido University,
Sapporo 060-8628, Japan}
\affiliation{Center of Education and Research for Topological Science and Technology, Hokkaido University,
Sapporo 060-8628, Japan}

\date{\today}% It is always \today, today,

\begin{abstract}
The Aharonov-Bohm (AB) effect has been accepted and has promoted interdisciplinary scientific activities in modern physics. To observe the AB effect in condensed matter physics, the whole system needs to maintain phase coherence, in a tiny ring of the diameter 1 $\mu$m and at low temperatures below 1 K. We report that AB oscillations have been measured at high temperature 79 K by use of charge-density wave (CDW) loops in TaS$_3$ ring crystals. CDW condensate maintained macroscopic quantum coherence, which extended over the ring circumference 85 $\mu$m. The periodicity of the oscillations is $h/2e$ in accuracy within a 10 \% range. The observation of the CDW AB effect implies Frohlich superconductivity in terms of macroscopic coherence and will provide a novel quantum interference device running at room temperature.
\end{abstract}

\maketitle

\section{Introduction}
50 years ago, Aharonov and Bohm challenged our understanding of vector potential that had not been a physical entity \cite{Aharonov}. Since then, the concept of AB (Aharonov and Bohm) phase has been accepted in modern physics \cite{Tonomura}, such as condensed matter physics \cite{Webb,Altshuler,AharonovC,Konig}, particle physics \cite{Wu}, non-abelian gauge theories \cite{Horvathy}, gravitational physics \cite{Ford,Ashtekar} and laser dynamics \cite{Hosten}. To observe AB phase in condensed matter physics, the whole system needs to maintain phase coherence, in a tiny ring of the diameter 1 $\mu$m and at low temperatures below 1 K, typically \cite{Webb}. We report first evidences for the AB interference effect at high temperature 79 K, by use of charge-density wave (CDW) loops in TaS$_3$ ring crystals \cite{Tanda} whose circumference is as long as 85 $\mu$m. CDW condensate maintained macroscopic quantum coherence which extended over the ring system \cite{Gruner}. Periodicity of the AB effect of CDW was half a flux quantum $h/2e$ with h and e being Plank constant and an elementary charge, respectively. Using interference of CDWs, a room temperature operating quantum interference device may be produced because some low dimensional materials, such as NbS$_3$, exhibit a CDW transition above room temperature. This will be substituted for superconducting quantum interference devices (SQUIDs), which can be only used at low temperatures.

CDWs have not been considered to exhibit AB effect because an electron-hole pair of CDW has neutral charge that cannot couple to vector potential, and the CDW phase is usually treated as a classical coordinate. However, the first prediction of AB effect of CDW was made by Bogachek {\it et al} \cite{Bogachek}. Their model includes a term representing the interaction of the CDW with the vector potential field, and then estimated the oscillation of $h/2e$. Another theory states that quasi-particle interference is generated by changing the CDW ground state \cite{Visscher,Yi,Montambaux}. Latyshev {\it et al} tried to observe the AB effect of CDWs in NbSe$_3$ crystals with tiny random holes pierced by heavy-ion irradiation \cite{Latyshev}. Oscillation in the nonlinear CDW conductivity as a function of the magnetic field was shown. Visscher {\it et al}, however, insisted that this could be the AB effect of a quasi-particle passing through a metallic region where the CDW order was destroyed near the tiny holes \cite{Visscher1998}. Therefore, existence of the AB effect of CDW condensate has been an open question as yet.

For this reason, a CDW loop, over which continuity of the CDW order is maintained, should be exploited in order to decide the argument. This cannot be realized simply by piercing a hole in a conventional single crystal, because continuous CDW chain doesn't be formed with the hole. Topological crystals (ring and M\"obius strip shaped seamless crystals) \cite{Tanda,Matsuura} are only materials providing such the CDW loop. The chain axis of CDW directs itself along the circumference of the ring without losing the CDW order, thereby forms a CDW loop.

Besides, the previous experiment of NbSe$_3$ had another problem. In the CDW state of NbSe$_3$, there still remain uncondensed electrons, which make it difficult to determine the origin of oscillation. Hence the choice of materials is crucial. We investigated orthorhombic TaS$_3$, which have a Peierls transition at 218 K, and no uncondensed electrons at low temperatures.

\section{Experimental}
The samples are synthesized by the chemical vapor transportation method \cite{Tanda}. The materials (tantalum and sulfur) in an evacuated quartz tube react to produce TaS$_3$ crystals (whisker, ring and other kinds of topological crystals). We prepared a ring/tube crystal reproducibly (Figure 1a), and then cut the one with focused ion beam (Figure 1c). The dimensions of the sample are 27 $\mu$m in diameter and 1 x 0.1 $\mu$m$^2$ in cross-section area. We fabricated gold electrode on the sample, by electron-beam lithography \cite{Inagaki}, after coating with polymethylmethacrylate (Figure 1b). 

We measured the electric current at 5.1 K while applying a magnetic field to the vertical the ring crystal (Figure 1b). The magnetic field is swept at a sweep rate of 50 mGauss/sec slow enough to eliminate generation of eddy current.

The magnetoresistance of a TaS$_3$ ring crystal is shown at the voltage, 100, 200 and 300 mV, in Figure 2(a). We observed periodic oscillations in these voltages, increasing the amplitude together with the increasing voltage. 

We obtained a peak $h/2e$ estimated by the power spectra of the observed oscillation. The insets of Figure 2(a) show the evolution of the power spectra, 100, 200 and 300 mV, from bottom to top. The period of the oscillation $\Delta B$ is 39.7 mGauss for all voltages. By assuming the origin of the oscillation to the AB effect \cite{Aharonov,Webb,Latyshev}, we estimated the effective charge $e^*$ using the following formula,
 
\begin{equation}
 \Delta B = \frac{\Phi}{S} = \frac{h}{e^*} \cdot \frac{1}{\pi r^2}
\end{equation}

where $h$ is Planck's constant and $r$ is the radius of the ring crystal. The estimated value of an effective charge e$^*$ is 3.0 x 10$^{-19}$ C within a 10 percent accuracy and is approximately equal to two times the elementary charge ($2e$ = 3.2 x 10$^{-19}$ C). The red lines in Figure 2(a) show the sinusoidal oscillation with the period and amplitude corresponding to the peak of the power spectra. The major contribution of the observed oscillations is well expressed by the obtained spectra. Other peak, such as $h/e$ or $h/4e$, was not observed within our experiment. 

We performed several experiments to confirm reproducibility of our reservation. Changing the sample temperature to 79 K did not show discrepancy in the oscillation period with data at 5.1 K (Figure 2(b)). The insets of Figure 2(b) show the evolution of the power spectra, 25, 50 and 75 mV, from bottom to top. The estimated value of an effective charge e$^*$ is 3.1 x 10$^{-19}$ C. Ordinarily, quantum phenomena hide in thermal fluctuation at high temperature by nature, but CDW state shows quantum interference at 79 K, such as superconductor. Moreover, another sample with diameter of 17 $\mu$m reproduced the same behavior with the period of 95.2 mGauss, which is corresponding to 3.1 x 10$^{-19}$ C. This periodic oscillation is also nearly equal to $2e$.

\section{Results and Discussion}

Figure 3 shows the voltage-current characteristics of the sample, after removing an Ohmic (linear) component of the conduction. The CDW current is shown to develop at larger voltage than the threshold $V_T$. The amplitude of the sinusoidal oscillation at $B$=0 was also plotted in Figure 3 (black line is guide for eyes in bottom panel). In the case of the ordinary fluctuation concerned with free electron, the line becomes linear as a function of the increasing voltage. Figure 3, however, shows that top and bottom curves behave similarly, in particular, a significant increase above $V_T$. For this reason, the amplitude of the interference is related with sliding CDW. It is plausible to attribute the development of the $h/2e$ oscillation to the CDWs.

In addition, the size of the sample is longer than coherent length of a quasi-particle. When a CDW was broken into quasi-particles, they can only stay within the coherence length of CDW. This is understood by analogy with a superconductor. Hence the quasi-particle will not travel over several nanometers, and the coherence of a quasi-particle is unlikely to be preserved over the whole system with the circumference of 85 $\mu$m at 79 K. The phase correlation length of sliding CDW is longer than quasi-particle and the circumference of our sample because sliding CDW becomes more ordered in the direction of motion \cite{Ringland}. In case of our sample, the possibility of A'tsuler-Aronov-Spivak effect \cite{Latyshev}, namely $h/2e$ oscillation of a quasi-particle, is ruled out. 

 Our study shows that the periodicity of CDW is half a flux quantum $h/2e$. Bogachek {\it et al} \cite{Bogachek} suggested that charge quanta of CDWs are multiple of chain number. For example, interference of a single CDW chain results in the period $h/2e$, while a bundle of CDWs with $N$ chains provides an oscillation of $h/2Ne$ period. If transverse correlation between CDWs formed CDW bundles, the period of $h/4e$ or $h/8e$ must be observed. In fact, the power spectra only give the fundamental $h/2e$ oscillation. 

Thus we propose a model to describe the $h/2e$ oscillation. Figure 4 shows a schematic picture of a CDW soliton confined into a single chain.  A CDW chain (the green wave in Fig. 4) has an extra wavefront, namely a soliton, carrying a charge of $2e$ \cite{Maki}. The yellow circle corresponds to a dislocation loop encircling the soliton \cite{Duan,Hatakenaka}. Existence of the dislocation loop makes the soliton to move freely along the chain accompanying the $2e$ charge. This agrees with our observation.
Finally, we estimated a mass of CDW condensate, which influences coherence and decoherence in quantum mechanics. The CDWs have large effective mass being 1000 times the electron mass per CDW wavelength as they coupled with phonons and lattice \cite{Gruner}. In the case of TaS$_3$ ring which has the circumference of 85 $\mu$m, the effective mass of the single CDW loop is estimated about 10$^8$ times electron mass \cite{Kagoshima}. On the other hand, if a CDW soliton is localized in space, the mass is comparable to the effective mass \cite{Arndt}. In this context, determination of the model for our observation will directly relate with an applicability of quantum mechanics on macroscopic scales, and solve an important open question in quantum mechanics.

In conclusion, we observed the AB effect at liquid-nitrogen temperature 79 K using a CDW loop of TaS$_3$ ring crystal. The observed period was $h/2e$, which cannot be explained by interference of a quasi-particle. A soliton confined in a CDW chain, possibly originated the $h/2e$ oscillation. This work suggests that macroscopic quantum tunneling \cite{Bardeen} and Fr\"ohlich superconductors \cite{Frohlich} can occur in sliding CDWs. In principle, the quantum interference of CDWs provides a same sensitivity to an applied magnetic field to that of superconducting quantum interference device. And note that the transition temperatures of CDWs are relatively higher than those of superconductor. Some materials show CDW behavior even at room temperature. Hence our observation will provide a Fr\"ohlich superconducting quantum interference device running at room temperature, which has not been achieved by superconductor.

\begin{acknowledgements}
The authors are grateful to T. Toshima, T. Tsuneta, T. Matsuura, K. Matsuda, K. Ichimura, and K. Yamaya, for experimental support and Y. Asano, M. Hayashi, and N. Hatakenaka for stimulating discussions. This research has been supported by Grant-in-Aid for the 21st Century COE program on gTopological Science and Technologyh and the Japan Society for the Promotion of Science.
\end{acknowledgements}

\newpage

\begin{figure}[h]
\begin{center}
\includegraphics[width=0.7\linewidth]{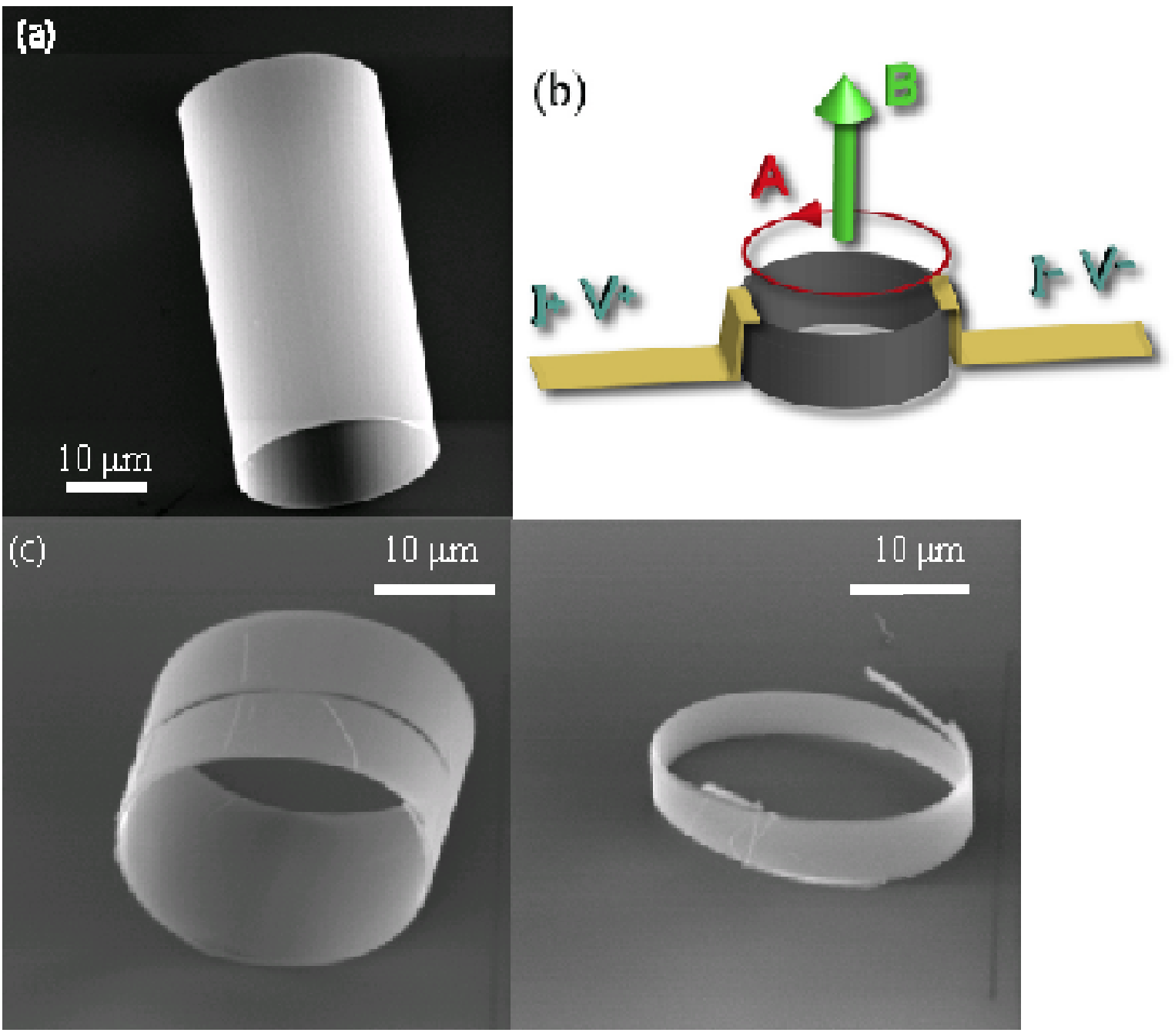}
\caption{(a) Scanning electron microscope image of a TaS$_3$ tube-shaped crystal. The diameter was about 27 $\mu$m and the thickness was about 0.1 $\mu$m. It is a seamless single crystal and the CDW chain direction coincides with the circumference direction. (b) Contact configuration for transport measurement. Two gold electrodes (100 nm thick, 5 $\mu$m wide) were evaporated. Electrical conduction measured at several constant voltages that changed when a magnetic field was applied to the vertical axis of a ring crystal. (c) Scanning electron microscope image of a TaS$_3$ tube/ring crystal. The tube crystal was cut by FIB. The lateral face of crystal was irradiated with ion beam, and then the crystal was pushed with a capillary tube. (Left) Before cutting, (right) after cutting.}
\label{fig_1}
\end{center}
\end{figure}

\newpage

\begin{figure}[h]
\begin{center}
\includegraphics[width=0.8\linewidth]{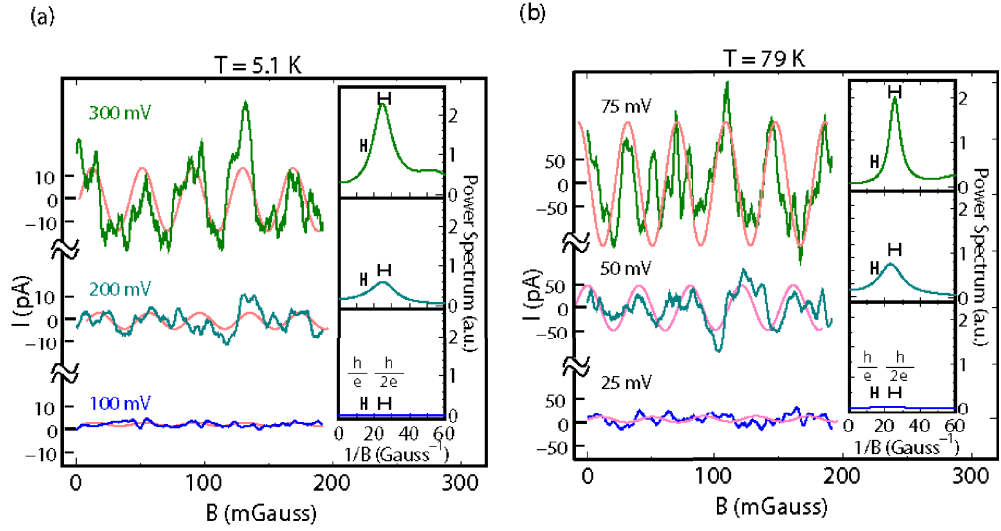}
\caption{(a) Change in current value as magnetic field is applied at 5.1 K. The period is observed at 100, 200 and 300 mV, (b) and 25, 50, and 75 mV at 79K. Insets show the power spectrum of the observed oscillation. As increasing the voltage over threshold voltage, the peak of the power spectrum gradually rise. The bars show that the periodic signal of $h/e$ and $h/2e$ is estimated by the area of the ring crystal. h and e is Plank's constant and elementary charge, respectively. The periods of the peak are nearly congruent with the one of $h/2e$. The red lines in Fig. 2 (a) and (b) show the sinusoidal oscillation with the period and amplitude corresponding to the peak of the power spectra. }
\label{fig_2}
\end{center}
\end{figure}

\newpage

\begin{figure}[h]
\begin{center}
\includegraphics[width=0.35\linewidth]{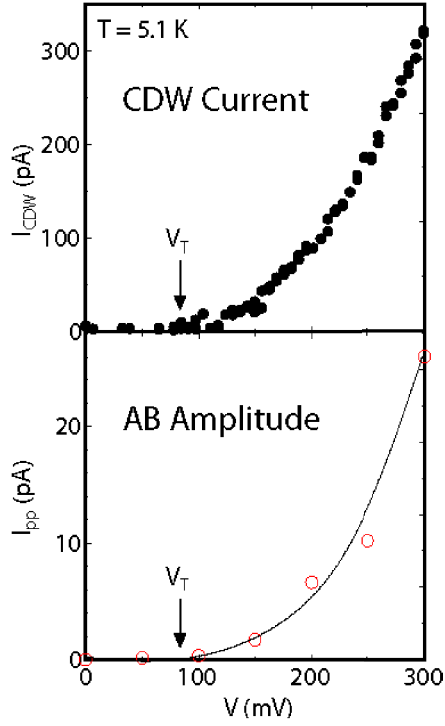}
\caption{(top) The voltage-current curve showing threshold voltage V$_T$ at 5.1 K. (bottom) The AB amplitude as a function of voltage. The black line shows guide for eyes. The AB amplitude means the amplitude estimated from the power spectrum. The curve of the amplitudes configurate with the voltage-current curve. For this reason, these phenomena derive from CDW origin. The parts of sliding CDW react as a function of magnetic field because Ipp is equal to 8 \% of I$_CDW$.}
\label{fig_3}
\end{center}
\end{figure}

\newpage

\begin{figure}[h]
\begin{center}
\includegraphics[width=0.7\linewidth]{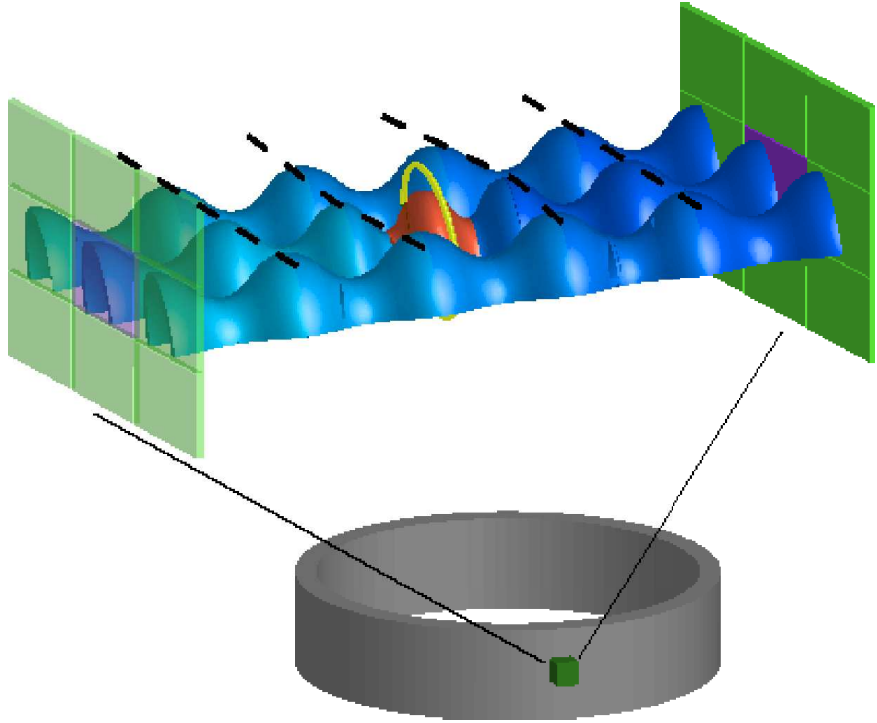}
\caption{A schematic picture of a CDW soliton encircled with a dislocation loop in the crystal. Each blue wave denotes the charge density of each chain, modulated with wavelength of $\lambda = 2\pi/2k_F$. In the central chain, an extra wavefront (red bump), namely CDW soliton, is encircled with a dislocation loop (yellow circle). The dotted lines show phase contours of CDW order parameter.}
\label{fig_4}
\end{center}
\end{figure}

\end{document}